# Co-axial dual-core resonant leaky fibre for optical amplifiers

**Ajeet Kumar[1], Vipul Rastogi[1*], Charu Kakkar[2] and Bernard Dussardier[3]**
[1]Department of Physics, Indian Institute of Technology, Roorkee 247 667, India
[2]Department of Physics, Kirorimal College, University of Delhi, Delhi 100 017 India
[3]Laboratoire de Physique de la Matière Condensée
  Université de Nice-Sophia Antipolis, Parc Valrose 06108 Nice, France

E-mail: vipul.rastogi@osamember.org

**Abstract.** We present a co-axial dual-core resonant leaky optical fibre design, in which the outer core is made highly leaky. A suitable choice of parameters can enable us to resonantly couple power from the inner core to the outer core. In a large-core fibre, such a resonant coupling can considerably increase the differential leakage loss between the fundamental and the higher order modes and can result in effective single-mode operation. In a small-core single-mode fibre, such a coupling can lead to sharp increase in the wavelength dependent leakage loss near the resonant wavelength and can be utilized for the suppression of amplified spontaneous emission and thereby gain equalization of an optical amplifier. We study the propagation characteristics of the fibre using the transfer matrix method and present an example of each, the large-mode-area design for high power amplifiers and the wavelength tunable leakage loss design for inherent gain equalization of optical amplifiers.

**Keywords:**  single-mode fibre, large-mode-area-fibre, leakage loss, confinement loss, fibre laser, optical communication, dense-wavelength-division-multiplexing, gain flattening, amplified spontaneous emission, erbium doped fibre amplifier

## 1. Introduction

In recent years there has been considerable research in designing optical fibres for applications, which include large-mode-area (LMA) operation and gain equalization of optical amplifiers [1-24]. LMA fibres are useful in building good beam quality high power fibre lasers and for high data rate dense wavelength division multiplexed optical fibre communication system. Kawakami et al. have used double clad fibre geometry to achieve ~ 2 times larger core area as compared to step index fibre [1]. With the recent advances in high power fibre lasers and need to scale core size for high power output has revived research in LMA fibres [3-8].  Different ways to design LMA fibres can be broadly categorized into i) scaling down the numerical aperture (NA) of the fibre to increase the core size, ii) use of photonic crystal fibres or holey fibres, and iii) use of special cladding geometries or index-profiles to discriminate higher order modes. Over the past few years, we have proposed some LMA designs using leaky fibres, where a large differential leakage loss between the fundamental and higher-order modes is responsible for single-mode (SM) operation [12-14, 21-23].  These designs include periodically arranged high- and low- refractive index segments in the angular direction of the cladding [12,21,22], a graded-index cladding with radially rising refractive index [13], and the cladding made of periodically arranged low-index trenches of varying strength in an otherwise high-index medium [14]. We have recently employed such a design in planar optical waveguides to achieve large-core SM operation [25].  In this paper we present an optical fibre design with co-axial dual cores and  leaky cladding. A resonant coupling of power from inner core to the highly leaky outer core makes the design efficient for achieving LMA SM operation. We analyze the fibre using the transfer matrix method (TMM) [26-27] and study the effects of various design parameters on the resonant coupling and the differential leakage loss between the fundamental and higher-order modes of a large inner core fibre to obtain an LMA design. We also analyze the wavelength dependence of the leakage loss of a small inner core SM fibre to obtain a design, which can suppress amplified spontaneous emission (ASE) and flatten the gain of an optical fibre amplifier in the desired wavelength band. Such a suppression of ASE in the

---
[*] Author to whom any correspondence should be addressed.

C-band of an erbium doped fibre amplifier (EDFA) helps in enhancing and flattening the gain in the S-band [20]. In particular, we present i) an LMA design with 30 μm core diameter and 0.16 NA capable of SM operation at 1550 nm wavelength by stripping off higher order modes after 36 cm propagation distance, and ii) a 5 μm core diameter design for C-band ASE suppression in an EDFA giving an inherent gain equalization around 20 dB in the S-band. The study should be useful in designing fibres for high power optical amplifiers and for gain equalization of optical fibre amplifiers.

## 2. Fibre design and method of analysis

The refractive index profile of the proposed fibre design is shown in figure 1 and is defined as

$$n(r) = \begin{cases} n_1; & 0 < r < a \\ n_2; & a < r < b \\ n_1; & b < r < c \\ n_3; & c < r < d \\ n_1; & r > d \end{cases} \quad (1)$$

where $n_1 > n_3 > n_2$.

The high-index regions $0 < r < a$ and $b < r < c$ define the inner and outer cores of the fibre respectively. The regions $a < r < b$ and $c < r < d$ are the depressed cladding regions and can be defined as inner and outer claddings respectively. The outermost high index region ($r > d$) makes the overall design leaky and all the modes suffer from finite leakage loss. Refractive index $n_2$ defines the NA of the fibre given by NA = $\boldsymbol{n_1}\sqrt{2\Delta}$, where $\boldsymbol{\Delta = \dfrac{n_1^2 - n_2^2}{2n_1^2}}$. Refractive index $n_3$ can primarily be used to achieve the resonance between a particular set of inner and outer core modes. To facilitate discussion we also define the widths of various layers as $d_1 = b\text{-}a$, $d_2 = c\text{-}b$, and $d_3 = d\text{-}c$.

We have used transfer matrix method (TMM) to calculate the leakage losses of the modes [26, 27]. TMM is particularly useful for analyzing multi-layer structure such as the one proposed in this paper. In TMM an arbitrary refractive index profile is divided into a large number of homogeneous layers by using the staircase approximation. By applying suitable boundary conditions at the interface of two consecutive layers, the field coefficients in the layers can be related by a 2 × 2 matrix. The field coefficients of the first and the last layer of the profile can then be connected by simply multiplying the transfer matrices of all the intermediate layers. By applying suitable boundary conditions in the first and the last layer, a complex eigenvalue equation is formed. The complex eigenvalue equation can then be solved to calculate the complex propagation constant by choosing a proper root searching algorithm. The effective index of a mode can be calculated from the real part of the propagation constant while the leakage loss can be estimated from the imaginary part.

In order to design the fibre, first we define the inner-core in terms of core-index $n_1$, core radius $a$, and the relative core-cladding index difference $\Delta$. For an LMA design, $a$ is large and in general supports several modes. We then add an outer core to the structure at $r = b$ of width $d_2$ (= $c\text{-}b$), core index $n_1$ and outer cladding index $n_3$. The outer core is separated from the inner core by the inner cladding layer of refractive index $n_2$ and width $d_2$. The entire structure is then made leaky by adding up a high-index layer of refractive index $n_1$ at $r = d$. The design parameters are so chosen that the $LP_{01}$ mode of the structure suffers from small leakage loss and the higher order modes have large leakage losses. Leakage losses of the modes in general increase with the mode order. However, by varying the outer cladding index $n_3$ there can be a resonant coupling of power from $LP_{11}$ mode to one of the modes of outer core, which is evidently highly leaky. This can increase the leakage loss of $LP_{11}$ mode significantly and in some cases it is even larger than the loss of $LP_{02}$ mode. Such a resonant coupling can, therefore, considerably increase the differential leakage loss between the fundamental and higher order modes and can result in an efficient LMA design.

For ASE suppression, the inner core is single-mode and the leakage loss of the mode in general increases with the wavelength due to the leaky cladding. However, a resonant coupling of power from the inner core to the highly leaky outer core can result in sharp increase of loss near the resonant wavelength, which can be tuned by varying the outer cladding parameters $n_3$ and $d_3$. Such a design can, thus, be used for example in ASE suppression and gain equalization of an S-band EDFA.

## 3. Numerical results and discussion

*3.1 LMA Design*
We define the inner core of the fibre by the following parameters:

$$n_1 = 1.444388, \Delta = 0.006, a = 15 \text{ μm} \tag{2}$$

The wavelength of operation used is 1.55 μm through out the paper, unless stated otherwise. We choose the outer core width $d_2 = 8$ μm and the separation between the two cores is $d_1 = 4$ μm. The high-index layers of the fibre can be made from pure silica. The low-index layers can be fabricated by fluorine doping in silica by modified chemical vapor deposition (MCVD) or by plasma activated chemical vapor deposition (PCVD) technique. MCVD can enable the value of Δ close to 0.5 % and with PCVD a value of Δ as large as 2 % can be achieved [28]. The refractive-index in the layers can be controlled by varying the concentration of fluorine. In our calculations we have used the value of Δ that should be possible to achieve by MCVD technique.
The fibre works on the principle of mode filtering. A high differential leakage loss between the fundamental and the higher order modes with a nominal loss to the fundamental mode ensures effective SM behavior of the fibre. We use outer cladding parameters $n_3$ and $d_3$ to primarily control the leakage losses of the modes. In figure 2 we have plotted the variation of leakage losses of the first three modes as a function of $n_3$. The leakage losses of the modes increase with the outer cladding index $n_3$. A peak appearing in the loss curve of $LP_{11}$ mode around $n_3 = 1.4425$ reflects the resonant coupling of power from $LP_{11}$ mode to the outer core, which is highly leaky. Such a coupling of power facilitates the stripping of $LP_{11}$ mode quickly, even before the $LP_{02}$ mode. At this resonance cladding index it is the differential leakage loss of $LP_{02}$ mode, which now determines the single-mode operation of the fibre. In this way the differential leakage loss between the fundamental and the higher order modes can be increased significantly and the minimum propagation length required for stripping off higher order modes by introducing 20dB or more loss can be brought down considerably. Next, we study the effect of outer cladding width $d_3$ on the leakage losses of the modes, which is shown in figure 3. The leakage loss of $LP_{01}$ mode significantly decreases with $d_3$. For higher order modes, one can note that in the range 2.4 μm < $d_3$ < 11.3 μm the leakage loss of $LP_{11}$ mode surpasses that of $LP_{02}$ mode and reflects resonant coupling effect. This range of $d_3$ can be used to design an LMA fibre. Figure 3 can, thus, help us in choosing the efficient design parameters. At the first crossing point ($d_3 = 1.4$ μm) the leakage loss of $LP_{11}$ and $LP_{02}$ modes is 186 dB/m and that of the $LP_{01}$ mode is 5 dB/m. In a design corresponding to this point, the higher order modes would strip-off in just 10 cm propagating distance. However, the fundamental mode would suffer from quite a large loss. When we increase the value of $d_3$ the differential leakage loss increases and also the loss of $LP_{01}$ mode goes down considerably. At the second crossing point ($d_3 = 11.3$ μm) the leakage loss of $LP_{11}$ and $LP_{02}$ modes is 55 dB/m and that of the fundamental mode is 0.04 dB/m. The higher order modes, thus, strip-off only after 36 cm propagation distance and the fundamental mode suffers from a nominal loss. The fibre, thus, shows an effective SM operation. The inner cladding thickness $d_1$ does not have significant effect on the differential leakage loss. This can however, be used to bring down the leakage losses of all the modes in the same proportion as can be seen in figure 4.
The spectral variation of leakage loss for the first three modes for a design corresponding to $d_3 = 10$ μm, which is close to the second crossing point of figure 3, is plotted in figure 5. The losses of the modes increase with wavelength due to more spread of the modal fields in the leaky cladding. The signature of resonance can be seen in crossing the loss curves of $LP_{11}$ and $LP_{02}$ modes at 1.48 μm wavelength, beyond which the leakage loss of $LP_{11}$ mode surpasses that of the $LP_{02}$ mode. However, due to the large dimensions of inner core the effect is not pronounced enough to result in a sharp variation of loss near the resonance wavelength.

The modal field patterns of various modes are shown in figure 6. The large leakage losses of $LP_{11}$ and $LP_{02}$ modes are reflected in relatively large amplitudes of oscillations in the cladding of the fibre. For $LP_{01}$ mode most of the power remains confined in the inner core and the spot size of the mode is 706 $\mu m^2$. An SM fibre with such a large mode area should be useful for high power lasers and amplifiers.

*3.2 S-band EDFA Design*

In an EDFA the gain in the S-band is usually suppressed because the ASE in the C-band builds up strongly due to its large gain coefficient, which depletes the pump. Therefore, in order to have sufficient gain in the S-band, it is necessary to suppress ASE buildup in the C-band. This requires a spectral variation of leakage loss such that the wavelengths longer than the signal band suffer a high differential leakage loss. It is further desirable that even among the signal wavelengths, the spectral variation of loss is such that it compensates for the spectral variation of gain coefficient and results in a spectrally flat gain.

In this section, we study the spectral variation of leakage loss of a structure with single-mode inner core and show the possibility of obtaining inherent gain flattening of an S-band EDFA. This requires low loss in the first half of the band and a sharp increase in the leakage loss in the second half. The leakage loss of the mode can be controlled by the outer cladding parameters $n_3$ and $d_3$. Figure 7 shows the spectral variation of leakage loss for the structure defined by $n_1 = 1.444388$, $\Delta = 1\%$, $a = 2.5$ $\mu m$, $d_1 = 9$ $\mu m$, $d_2 = 2$ $\mu m$ and $d_3 = 10$ $\mu m$ for three different values of $n_3$. We may note that for $n_3 = 1.4330$ the loss is small in the entire band and for $n_3 = 1.4340$ the loss is very large even for shorter wavelengths. The loss spectrum corresponding to $n_3 = 1.4335$ meets the desired variation needed for gain equalization of an S-band EDFA. Next, we study the effect of outer cladding width $d_3$ on the loss spectrum as shown in figure 8, where we have plotted the leakage loss as a function of wavelength corresponding to $n_3 = 1.4335$ and three different values of $d_3$. One may note that for a given value of $n_3$, $d_3 = 8$ $\mu m$ causes a large leakage loss for all the wavelengths, while $d_3 = 12$ $\mu m$ results in very small loss. The loss spectrum corresponding to $d_3 = 10$ $\mu m$ shows a sharp increase in the leakage loss for wavelengths longer than 1.52 $\mu m$, which can suppress amplification of spontaneous emission in the C-band of the EDFA and can shift the gain to the S-band, where the leakage loss is quite small. Such a tunable leakage loss has been employed by Kakkar et al. [20] in a segmented cladding design. We show that the tunable leakage loss obtained in the proposed dual-core leaky design can suppress the ASE effectively with less stringent design parameters due to a completely leaky structure. We have carried out amplifier gain calculations using the model of Pedersen [29] for a fiber with the inner core doped with $Er^{3+}$ ions at a concentration of $5 \times 10^{24}$ $m^{-3}$. The gain spectrum has been simulated for 16 signal wavelengths between 1490 and 1520 nm propagating simultaneously in the fibre. Each of the signal wavelengths carried an input power of 8 $\mu W$ and the input pump power was 200 mW at 980 nm wavelength. The gain spectrum of the amplifier is shown in figure 9, where the dashed curve corresponds to the perfectly guided structure with fibre length $L = 8$ m, optimum for maximum gain, and the solid curve corresponds to the co-axial dual-core leaky structure with $L = 16$ m. The gain spectrum clearly shows the suppression of ASE in the C-band and shifting of gain to the S-band of the amplifier, thereby an inherent equalization of gain around 20 dB. Such an S-band EDFA leaky design should be possible to fabricate in silica glass with fluorine doping and should be useful for gain equalization of optical fibre amplifiers.

**4. Conclusions**

We have proposed a dual concentric cores leaky optical fibre design showing resonant coupling of power between the two cores. The design has been made for an LMA fibre in fluorine doped silica with 0.16 NA and 706 $\mu m^2$ spot-size. The fibre strips off higher order modes after 36 cm propagation and shows effective SM operation. The wavelength tunable leakage loss of the proposed structure has been utilized to make a design for ASE suppression of an EDFA and thereby gain enhancement in the S-band. With such a design we show a 20 dB flat gain in the S-band EDFA. The designs should be useful for optical fibre amplifiers and lasers for high power applications as well as for gain flattening.

**Acknowledgement**

This work has been partially supported by the Indo-French Networking Project on "R &D on specialty optical fibres and fibre-based components for optical communication."

**Figure Captions**

**Figure 1.** Schematic of the refractive index profile of a co-axial dual-core leaky fibre.

**Figure 2.** Variations of the leakage losses of the first three modes of the fibre with outer cladding index $n_3$ for outer cladding width $d_3 = 4$ μm.

**Figure 3.** Variations of the leakage losses of the first three modes of fibre with outer cladding width $d_3$ for outer cladding index $n_3 = 1.4425$. One can see two crossing points for $LP_{11}$ and $LP_{02}$ modes at $d_3 = 2.4$ μm and $d_3 = 11.3$ μm.

**Figure 4.** Effect of inner cladding width $d_1$ on the leakage losses of the modes for $d_3 = 10$ μm (near the second crossing point of figure 3).

**Figure 5.** Spectral variation of leakage losses of the first three modes of the fibre for $d_3 = 10$ μm. The crossing of the curves for $LP_{11}$ and $LP_{02}$ modes occurs at 1.48 μm wavelength.

**Figure 6.** Normalized modal field amplitudes for the first three modes of the fibre at 1.55 μm wavelength for $d_3 = 10$ μm.

**Figure 7.** Spectral variation of leakage loss for a fibre with $a = 2.5$ μm, $\Delta = 1\%$, $d_1 = 9$ μm, $d_2 = 2$ μm, $d_3 = 10$ μm and for three different values of $n_3$.

**Figure 8.** Spectral variation of leakage loss for a fibre with $a = 2.5$ μm, $\Delta = 1\%$, $d_1 = 9$ μm, $d_2 = 2$ μm, $n_3 = 1.4335$ and for three different values of $d_3$.

**Figure 9.** Gain spectrum of the fibre corresponding to $n_3 = 1.4335$, $d_3 = 10$ μm and all the other parameters corresponding to figure 7.

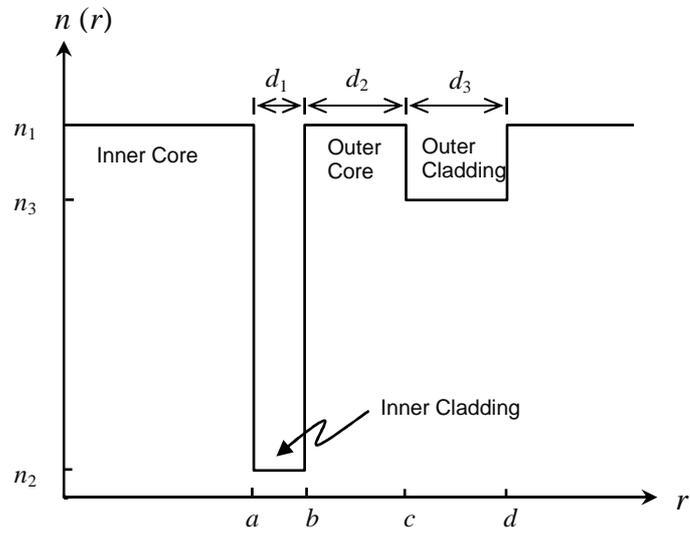

Figure 1

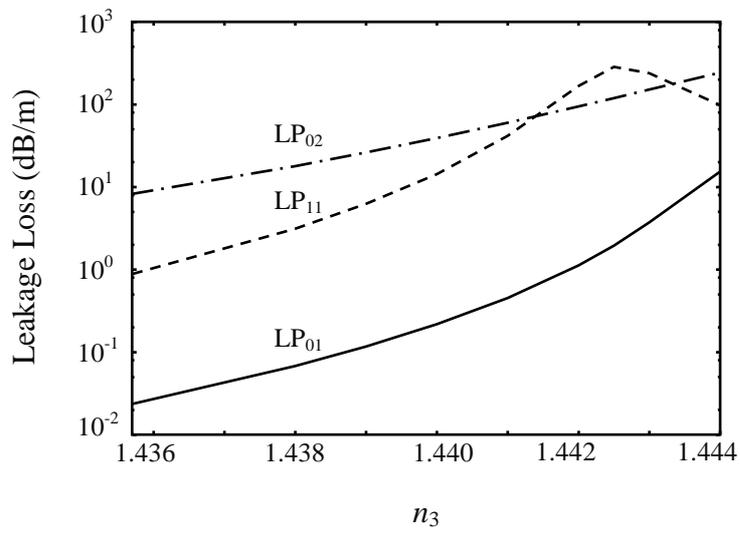

Figure 2

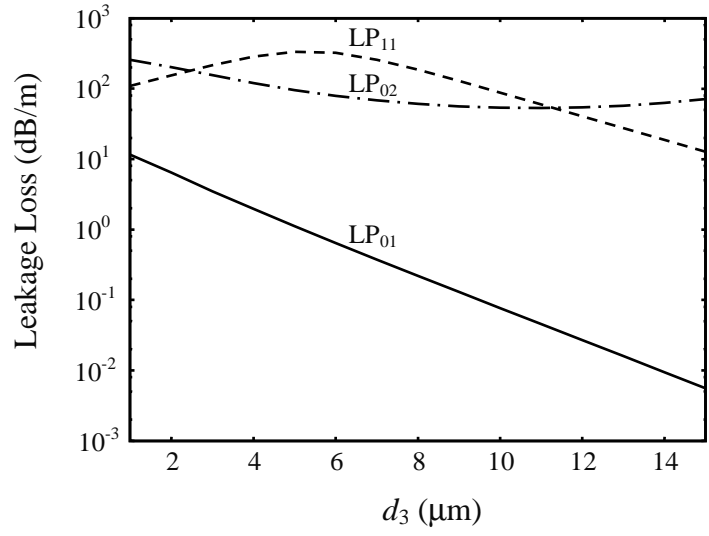

Figure 3

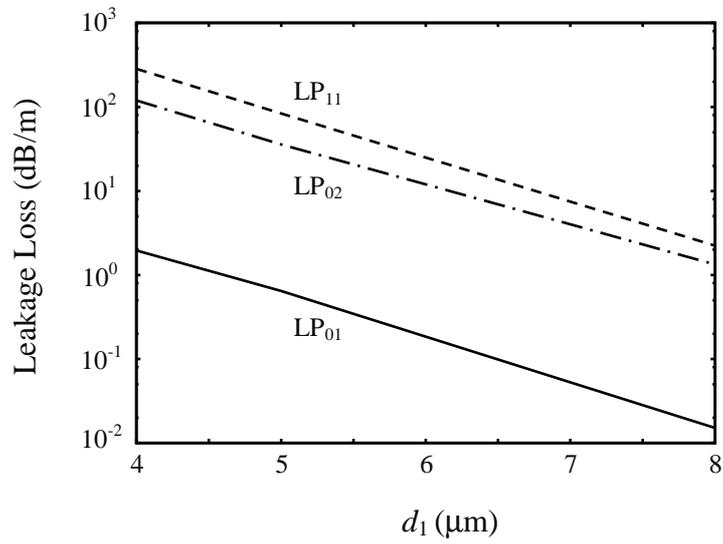

Figure 4

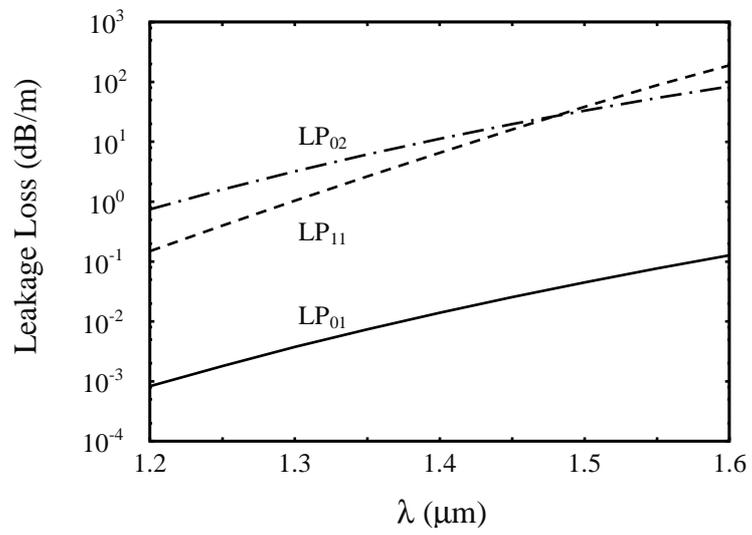

Figure 5

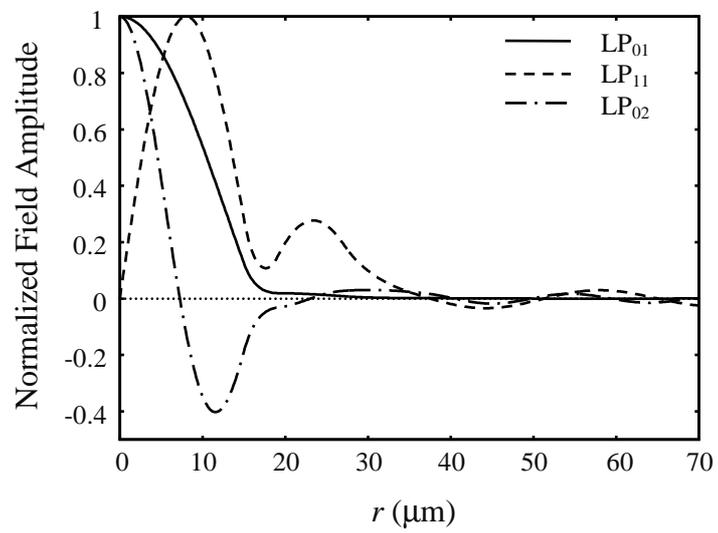

Figure 6

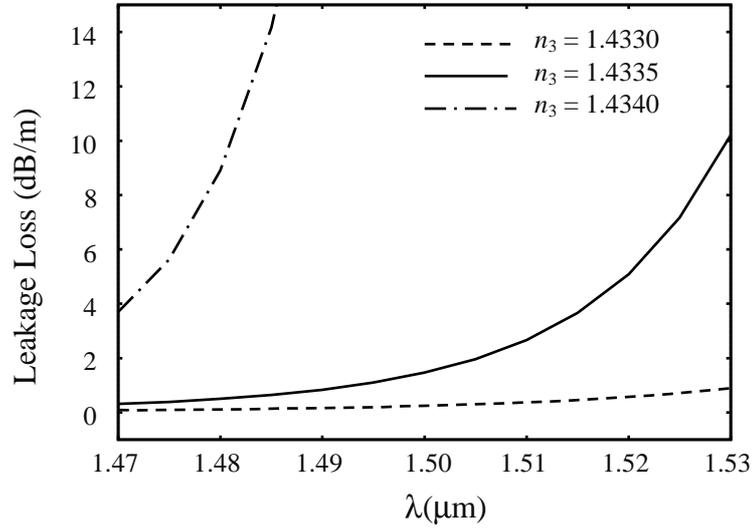

Figure 7

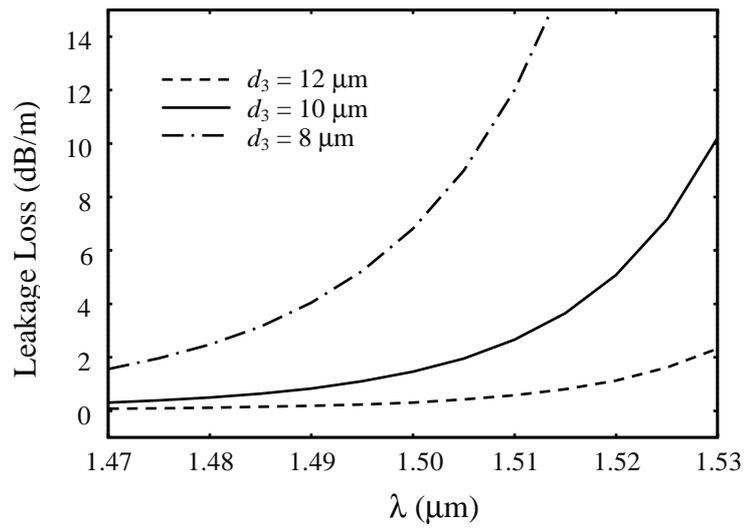

Figure 8

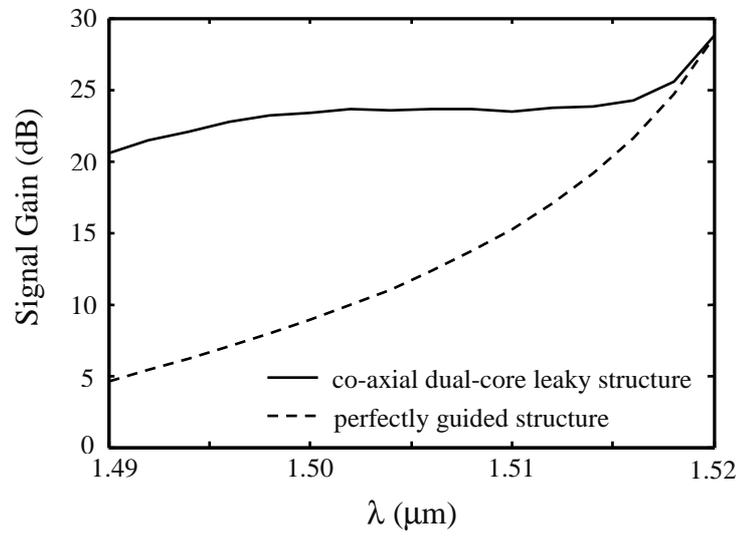

Figure 9